\documentclass[
 reprint,
 preprintnumbers,
 amsmath,
 amssymb,
 aps,
 prstab,
 floatfix,
]{revtex4-1}

\usepackage{graphicx}			
\usepackage{verbatim}			
\usepackage{wasysym}			
\usepackage{mathtools}			
\usepackage[makeroom]{cancel}		
\usepackage{dcolumn}			
\usepackage{hyperref}			
\usepackage{color}			

\newcommand{\pd}{\partial}				
\newcommand{\dd}{\mathrm{d}}				

\newcommand{\Nb}{N_{\text{p}}}		
\newcommand{\Qb}{ Q_{\beta}}		
\newcommand{\Qs}{ Q_s}		

\begin{document}


\title{Transverse Instabilities of a Bunch with Space Charge, Wake and Feedback}

\author{Alexey Burov}
\email{burov@fnal.gov}
\affiliation{Fermilab, PO Box 500, Batavia, IL 60510-5011}
\date{\today}

\begin{abstract}
When a resistive feedback and single-bunch wake act together, it is known that some head-tail modes may become unstable even without space charge. This {\bf f}eedback-{\bf w}ake {\bf i}nstability, {\sf FWI}, modified by space charge to a certain degree, is shown to have a special single-maximum increasing-dropping pattern with respect to the gain. Also, at sufficiently large Coulomb and wake {\bf f}ields, as well as the {\bf f}eedback gain, a new type of transverse {\bf m}ode-{\bf c}oupling {\bf i}nstability is shown to take place, {\sf 3FMCI}, when head-to-tail amplified positive modes couple and the growth rate saturates with the gain.
\end{abstract}


\maketitle

 


\section{Introduction}

For a bunched beam in a circular machine, the {\bf t}ransverse {\bf m}ode {\bf c}oupling {\bf i}nstability, or {\sf TMCI}, is the minimal one: it may happen when all potential sources of bunch instabilities, except single-bunch wakes, do not play a role. Even without Landau damping, this instability has a threshold wake amplitude, or, better to say, a threshold coherent tune shift proportional to the synchrotron frequency. For hadron beams, however, the wake fields are almost always accompanied by non-negligible Coulomb fields of their space charge ({\sf SC}), so a question of possible modification of TMCI by SC for such beams is of a primary importance. For the last twenty years, this problem has been addressed theoretically by several authors, see e.g.~\cite{blaskiewicz1998fast, Ng:1999fy, burov2009head, PhysRevSTAB.14.094401, Zolkin:2017sdv}. In particular, it has been shown that the TMCI wake threshold typically increases with the SC tune shift; at sufficiently strong SC, the threshold value of the wake-related coherent tune shift $w_\mathrm{th}$ is about the same as the SC tune shift $q$,
\begin{equation}
w_\mathrm{th} \simeq q \,.
\label{WthSC}
\end{equation}
This border of stability has an intriguing feature: it does not depend on the bunch population; thus, being satisfied for some number of particles, it will be satisfied for any number of them larger by any amount. This counterintuitive conclusion apparently contradicts the logic of the fast microwave instability, according to which at a sufficiently large number of particles, the instability should develop within a bunch similarly to the equivalent coasting beam, i.e. the coasting beam of the same line density and rms momentum spread~\cite{Ruth:1981xt}. For a coasting beam, though, the wake instability threshold drops exponentially with SC: the latter separates coherent tunes from the incoherent spectrum and thus suppresses Landau damping~\cite{PhysRevSTAB.12.034201}. Thus, either Eq.~(\ref{WthSC}) or the conventional microwave idea is incorrect, or, maybe, something important is missing in this picture. A hint to a resolution of these contradictions could be seen in Ref.~\cite{Cappi:2000ze}, where the fast instability was treated as beam breakup, thus yielding an estimation for head-to-tail amplification, not a growth rate all-over the bunch. In fact, this view had nothing to do with mode coupling, beside a similarity of the estimating formulas for the no-SC TMCI threshold and limiting beam breakup; this similarity actually played a misleading role. This beam breakup approach was left logically unfinished for many years; the estimated amplification along the bunch was misunderstood as 'TMCI' not only in Ref.~\cite{Cappi:2000ze}, but also in many later publications, notwithstanding the fact that mode coupling was never demonstrated for such strong space charge, as it was under consideration.  A resolution of the conundrum was recently suggested in Refs.~\cite{Burov:2018pjl, Burov:2018rmx}: while SC indeed makes the TMCI threshold {\it vanish}, it brings into the same place other types of instabilities, {\it convective} ones, with their derivatives, {\it core-halo instabilities} and {\it absolute-convective instabilities}. 

The stability problem for bunched beams with SC should not be considered solved unless such factors as the coupled-bunch interaction, chromaticity, feedback and Landau damping are taken into consideration as they are for no SC case~\cite{chao1993physics, PhysRevSTAB.17.021007}. In this paper it will be shown how the first three of these factors can be, in general, effectively taken into account, focusing, though, on the variety of solutions with the conventional resistive feedback. It will be shown that such dampers may suppress TMCI and may drive other instabilities, named here the {\bf f}eedback-{\bf w}ake {\bf i}nstability, {\sf FWI}, and the {\bf 3}-{\bf f}actor {\bf m}ode-{\bf c}oupling {\bf i}nstability, {\sf 3FMCI}. 

Our main instrument here is the same as in Refs.~\cite{Burov:2018pjl, Burov:2018rmx}: the {\bf a}ir-{\bf b}ag {\bf s}quare well, or {\sf ABS} model, suggested twenty years ago by M.~Blaskiewicz~\cite{blaskiewicz1998fast}. In its basic form, the model considers the bunch particles having the same longitudinal action within a square potential well; the particles interact with each other through wake and SC fields. This paper continues uncovering possibilities of the ABS model, showing how feedbacks may change beam dynamics in the presence of wakes and SC.

\section{ ABS Model} \label{Sec:ABS}

Starting from the case of a single bunch and a short-range wake (i.e. no multi-turn wakes), the original ABS equations of motion~\cite{blaskiewicz1998fast} can be presented in the form of Ref.~\cite{Burov:2018pjl}: 
\begin{equation}
\frac{\pd\,x}{\pd \theta} + \frac{\pd\,x}{\pd \psi} =  i q (x - \bar{x}) + i\,F\,,
\label{MainX0}
\end{equation}
with
\begin{equation}
\begin{split}
& F(\psi) = w \int_0^{|\psi|} \frac{ \dd \psi'}{\pi} W(\psi'/\pi) \bar{x}(\psi-\psi') \, ;\\
& \bar{x}(\psi) \equiv x(\psi)/2 + x(-\psi)/2 \,.\\
\end{split}
\label{MainX2}
\end{equation}
Here $x=x(\theta,\psi)$ is a slow amplitude of the transverse oscillations as a function of time $\theta$ and the synchrotron phase $\psi$, both measured in the synchrotron radians; the searched-for function $x$ is periodical on the phase $\psi$, which is supposed to change from $-\pi$ to $0$ for the tail-to-head moving particles, or for the $+$ flux, and from $0$ to $\pi$ for the $-$ flux, which runs back to the tail; the bunch length is 1 in these units. The term $\propto q$ is the SC force, and $F$ is the wake force with $W(s)$ as a dimensionless wake function; $\bar{x}(\psi)$ is the local centroid. The dimensionless wake parameter $w$ keeps in itself all dimensional values of the problem: 
\begin{equation}
\label{WParam}
	w = \frac{\Nb W_0 r_0 R_0}{4\,\pi\,\gamma\,\beta^2\,\Qb \Qs}\,,
\end{equation}
with $\Nb$ as the number of particles in the bunch, $W_0$ as the amplitude of the wake function in conventional units of Ref~\cite{chao1993physics}, $r_0$ as the particle classical radius, $R_0$ as the average radius of the machine, $\gamma$ and $\beta$ as the relativistic factors, $\Qb$ and $\Qs$ as the betatron and synchrotron tunes. 

Unless the bunch is very long, a typical feedback reacts on its center of mass ({\sf CM}) only, kicking the bunch as a whole; in this case the feedback term can be just added to the right-hand side of Eq.~(\ref{MainX0}) as $g_f \int_0^\pi \dd \psi \, \bar{x}/\pi$, where $g_\mathrm{f}$ is the complex gain. In this convention, the so called resistive damper corresponds to a pure real and negative gain. Except for a special case of long-range high-frequency wakes, the coupled-bunch interaction mostly reduces to a cross-talk between their centers of mass; thus, it is described similarly to the CM feedback, with the complex growth rate of the most unstable coupled-bunch mode $g_\mathrm{cb}$ as the gain; the latter can be found within a simple model of bunches as macroparticles, see e.g. Ref.~\cite{PhysRevSTAB.17.021007}. 

The chromaticity describes dependence of the particle transverse, or betatron, frequency on its momentum. It can be taken into account by addition of $\,\,i \zeta\, \mathrm{sgn}(\psi) \,x$ to the right-hand side of Eq.~(\ref{MainX0}), with $\zeta$ as the chromaticity parameter and $\mathrm{sgn}$ as the sign-function, yielding
\begin{equation}
\frac{\pd\,x}{\pd \theta} + \frac{\pd\,x}{\pd \psi} =  i q (x - \bar{x}) + i\,F\, +i \zeta \mathrm{sgn}(\psi) x + g \int_0^\pi \frac{\dd \psi}{\pi} \, \bar{x} ,
\label{MainX1}
\end{equation}
where the full CM gain $g=g_\mathrm{f}+g_\mathrm{cb}\,$. The chromatic term can be better dealt with the substitution $x \rightarrow x \exp(i \zeta |\psi|)$, and the sought-for offset $x$ can be Fourier-expanded over the synchrotron phase $\psi$, $x=\sum_n A_n \exp(i n \psi)$, leading to the following set of equations for the Fourier amplitudes:   
\begin{equation}
i\dot{A}_n = n A_n - q(A_n - \bar{A}_n) + \sum_{m=-\infty}^{\infty} (ig D_{nm}-w U_{nm} )\bar{A}_m\,.
\label{Aeq0}
\end{equation}
Here, the centroid Fourier amplitude $\bar{A}_m = (A_m + A_{-m})/2$, and the dimensionless wake and damper matrix elements are 
\begin{equation}
\begin{split}
& U_{nm} \equiv  \! {\int_0^1\! \! \! \dd s \int_0^s \! \! \! \dd s' W(s-s') e^{-i \zeta(s-s')} \cos(\pi ns) \cos(\pi m s')}\,; \\
& D_{nm} \equiv \frac{1- (-1)^{n} e^{-i\pi \zeta}}{\zeta^2-n^2} \,\frac{1- (-1)^{m} e^{i\pi \zeta}}{\zeta^2-m^2}\, \frac{\zeta^2}{\pi^2}\,.
\end{split}
\label{UnmDnm}
\end{equation}
The diagonal matrix elements of the damper matrix $D$ describe how the CM offset is distributed over the harmonics at the given chromaticity; that is why $\sum_{n=-\infty}^{\infty}D_{nn}=1$ for any $\zeta$. For purposes of this paper, the wake matrix elements $U_{nm}$ are computed with the Heaviside theta-function $W(s)=\Theta(s)$ everywhere except Sec.~\ref{Sec:ShortWake}, where a short broadband wake is examined.

\section{Two-Factor Instabilities}

While suppressing some modes, feedbacks may contribute to growth rates of other ones~\cite{PhysRevSTAB.17.021007, PhysRevAccelBeams.19.084402, Metral:2018lds}. In any case, feedbacks make some modes less stable, than they would be otherwise. Benefits from their usage are due to the fact that the growth rates which they contribute are normally more tolerable than those they suppress; so the relatively small remnant instabilities, either contributed by the dampers, or not fully eliminated by them, can be successfully overcome by Landau damping. The other part of the picture relates to SC, which makes bunch slices more rigid, and thus more coupled with the wake. Thanks to that, SC introduces head-to-tail amplification to the positive modes; for strong SC, the amplification can be large~\cite{Burov:2018pjl}. Due to this circumstance, in turn, the growth rates contributed by a damper may shoot up sharply.
\begin{figure}[h]
\includegraphics[width=\linewidth]{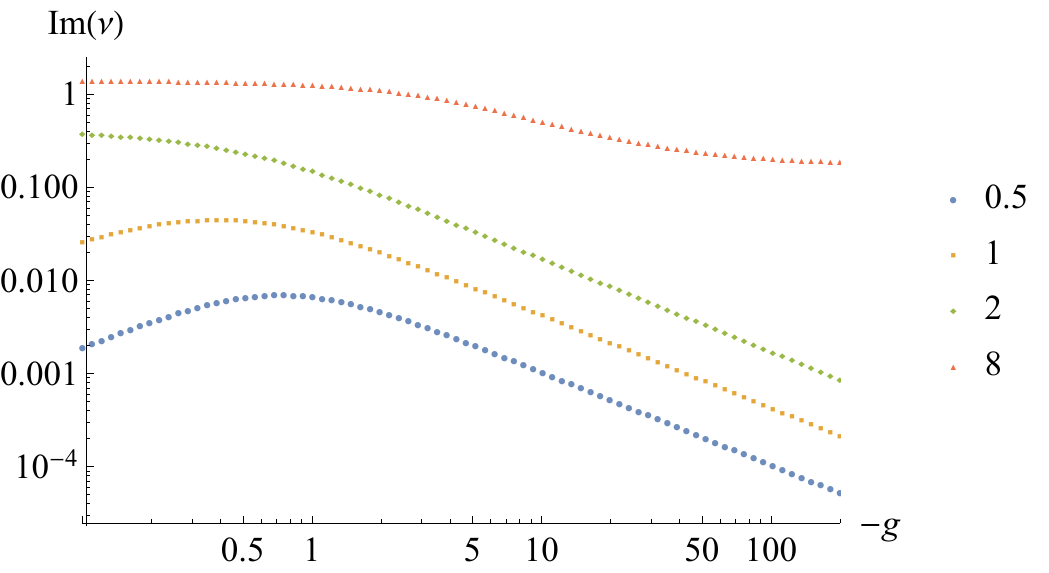}
\caption{\label{RatesVsGQ0}
	Growth rates of the most unstable modes (MUM) versus damping gain at zero SC; the legend shows values of the wake parameter $w$. The conventional TMCI wake threshold is $w_\mathrm{th}^0=1.15$, which explains opposite derivatives of $w=1$ and $w=2$ curves at zero gain. Below the TMCI threshold, $\Im \nu \propto |g|$ at $|g| \ll 1$. In the opposite case of $|g| \gg 1$, $\, \Im \nu \propto 1/|g|$ for wake below the high-order mode coupling, $w \leq 7$ in this case.}
\end{figure}

Before we start the exploration of the rather complicated case when SC, wake and damper are all present, let us recall how any two of these three factors work together at zero chromaticity. 
\begin{itemize}
\item{ SC $+$ Wake, no Feedback: As already mentioned, SC suppresses TMCI of the negative modes~\cite{Zolkin:2017sdv}, and causes or boosts convective instabilities of positive modes~\cite{Burov:2018pjl}. }
\item{ SC $+$ Feedback, no Wake: the feedback damping rate is distributed between the head-tail modes in one or another way, depending on the SC parameter; in any case, each mode remains stable~\cite{Burov:2016jsh}. }
\item{ Wake $+$ Feedback, no SC~\cite{PhysRevSTAB.17.021007, PhysRevAccelBeams.19.084402, Metral:2018lds}: the case is presented in Fig.~\ref{RatesVsGQ0}. At small gains and below the TMCI threshold, $|g| \ll1\,$ and $ w<w_\mathrm{th}^0$, the most unstable mode (MUM), which is the mode $-1$ here, gets a growth rate $\Im \nu \propto |g|$; for $w \ll w_\mathrm{th}^0$, $\; \Im \nu = 4\,|g|\, w^2/\pi^4$ for the theta-wake. This instability is caused by a positive feedback which a non-zero mode may receive from a resistive damper at non-zero wake; so it may be called the {\bf f}eedback-{\bf w}ake {\bf i}nstability, or {\sf FWI}. At a sufficiently large gain, the FWI growth rate drops with it, $\Im \nu \propto 1/|g|$. Thus, FWI has a single-maximum pattern in its dependence on the gain, being proportional at small gains and inversely proportional at large ones. For higher-order mode coupling, the CM damper does not help much, as seen for $\,w=8$ of Fig.~\ref{RatesVsGQ0}. The eigenvector analysis shows that for large gains, two pairs of modes couple there, $-1$ and $-2$, as well as $+1$ and $+2$; the most unstable mode is of the negative pair.}
\end{itemize}

\begin{figure}[h!]
\includegraphics[width=\linewidth]{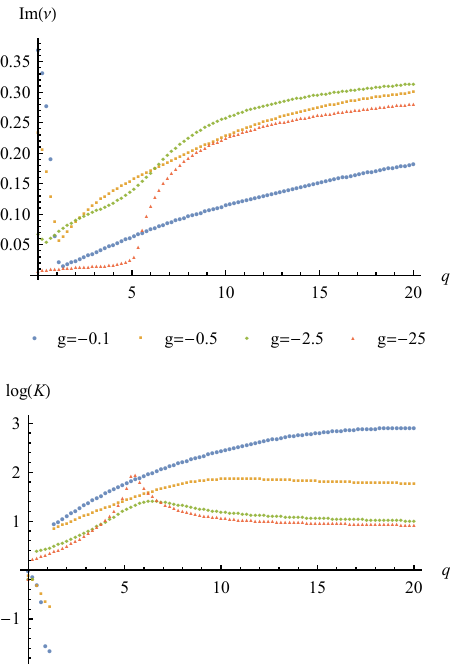}
\caption{\label{SCScan}
	Growth rate (top) and natural logarithm of amplification (bottom) versus SC parameter $q$ for the most unstable mode at the wake amplitude $w=2$, almost two times above the no-SC TMCI threshold, and various feedback gains $g$, shown in the legend. TMCI at small $q$ is seen to vanish when SC or gain increase. Larger gain suppresses the instability at $q \leq 5$; above that, $+1^\mathrm{st}$ and $+2^\mathrm{nd}$ modes couple (see Figs.~\ref{IDCM}, \ref{GridXSCGCoupled} below for more details), resulting in an almost gain-independent growth rate.      
	}
\end{figure}
\begin{figure}[h!]
\includegraphics[width=\linewidth]{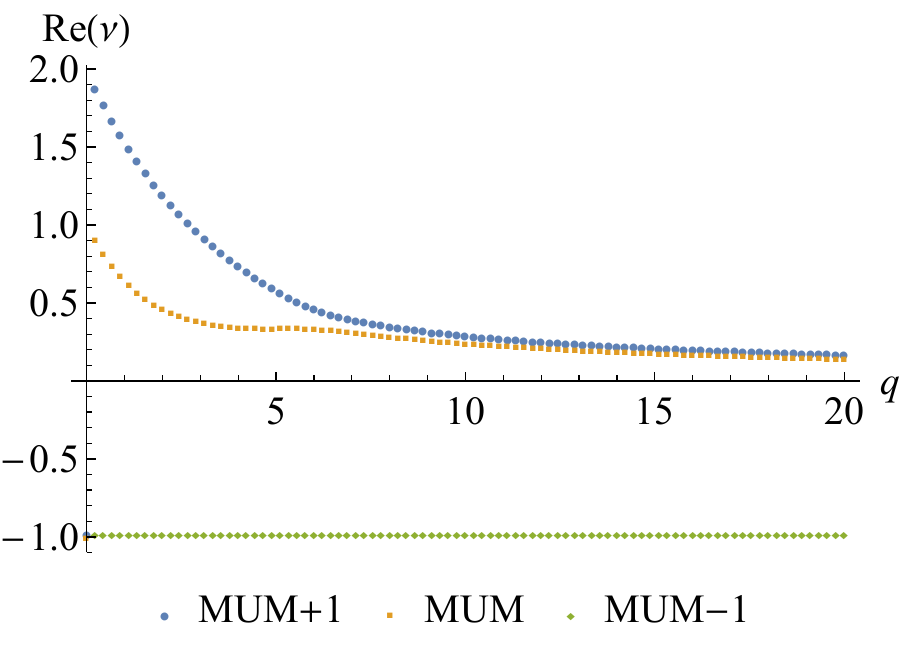}
\caption{\label{IDCM}
	With sufficient gain and wake, SC becomes a mode coupling factor, as in here, with $+1^\mathrm{st}$ and $+2^\mathrm{nd}$ modes coupling. The wake is same, $w=2$; the gain $g=5$.  MUM is $+1^\mathrm{st}$. Mode $0$ (green dots) is insensitive to SC, being just shifted down by $w/2=1$.}
\end{figure}
\begin{figure*}
\includegraphics[width=\linewidth]{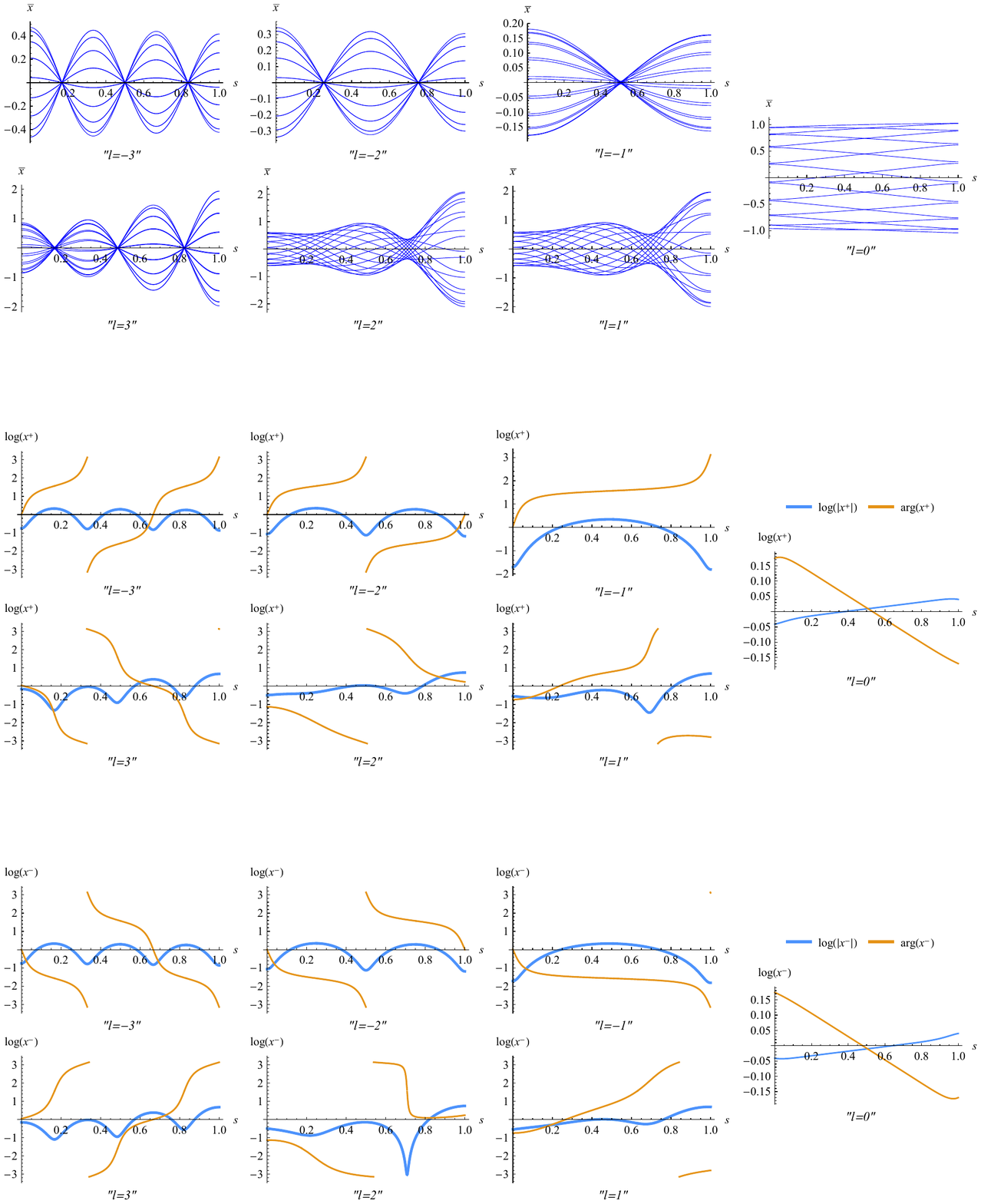}
\caption{\label{GridXSCGCoupled}
Eigenfunctions at $w=2,\, q=8,\, g=5$: centroid stroboscopic images (two top rows), tail-to-head "$+$" flux amplitudes (two middle rows) and head-to-tail "$-$" flux amplitudes (two bottom rows). Modes $+1^\mathrm{st}$ and $+2^\mathrm{nd}$ are coupled due to common action of the {\bf{F}}eedback, Coulomb {\bf{F}}ield  and wake {\bf{F}}ield, thus demonstrating {\bf 3}-{\bf f}actor {\bf m}ode-{\bf c}oupling {\bf i}nstability, {\sf 3FMCI}. Note their waists instead of nodes in the centroid stroboscopic and running phases (traveling wave) patterns for the $+$ and $-$ fluxes.
	}
\end{figure*}

\section{Three-Factor Mode-Coupling Instability}

Let us consider now the more complicated case with SC, wake and the resistive damper all involved, and see what sort of instabilities may be there. Figure~\ref{SCScan} shows how the growth rate of the most unstable mode (MUM) $\Im \nu$ (top) and natural logarithm of its head-to-tail amplification $\log K$ (bottom) change with SC at wake parameter $w=2$, which is almost double that of the no-SC no-gain TMCI threshold, for various damper gains $g$. The MUM is $-1^\mathrm{st}$ for a few points at small SC, and $+1^\mathrm{st}$ for the rest of the points; this switch of the MUM number can be recognized on both plots by a change in their behavior at small $q$. This switch can be expected: the mode $-1$ wins at low SC, since it is coupled with $0$ mode at no-gain case, while the mode $+1$ is selected by the absolute-convective instability, ACI, for higher SC, since this mode is positive, i.e. coupled with wake, and the resistive damper works as a positive feedback for it. Several things are worth noting with respect to Fig.~\ref{SCScan} and elucidating its features Figs.~\ref{IDCM} and \ref{GridXSCGCoupled}: 
\begin{itemize}
\item{TMCI at small SC and low gain is fully suppressed when either SC or gain, or both, become larger. It is a case when SC can make the bunch more stable.}
\item{At a sufficiently high gain, there is a pronounced SC threshold of the instability: it is deeply suppressed below that value, $q \approx 5$ here, and appears as TMCI above the threshold. This mode-coupling instability requires significant wake, SC and gain; each of the three factors has to work. Thus, it can be called {\bf 3}-{\bf f}actor {\bf m}ode-{\bf c}oupling {\bf i}nstability, {\sf 3FMCI}, which can also be read as {\sf{F}}eedback, Coulomb {\sf{F}}ield  and wake {\sf{F}}ield {\sf MCI}. }
\item{Above the 3FMCI threshold, the growth rate quickly becomes close to its gain-asymptotic value.}
\item{Eigenfunctions above the 3FMCI threshold are shown in Fig.~\ref{IDCM}, demonstrating typical TMCI features of the coupled and uncoupled modes: waists instead of nodes for the coupled modes, whose centroid patterns are almost identical and phases rather run than stand.}
\item{ There is a feature of the 3FMCI modes, which clearly differentiate them from the conventional TMCI ones: the coupled modes are positive, and thus convectively amplified.}
\end{itemize}
\begin{figure}[h]
\includegraphics[width=\linewidth]{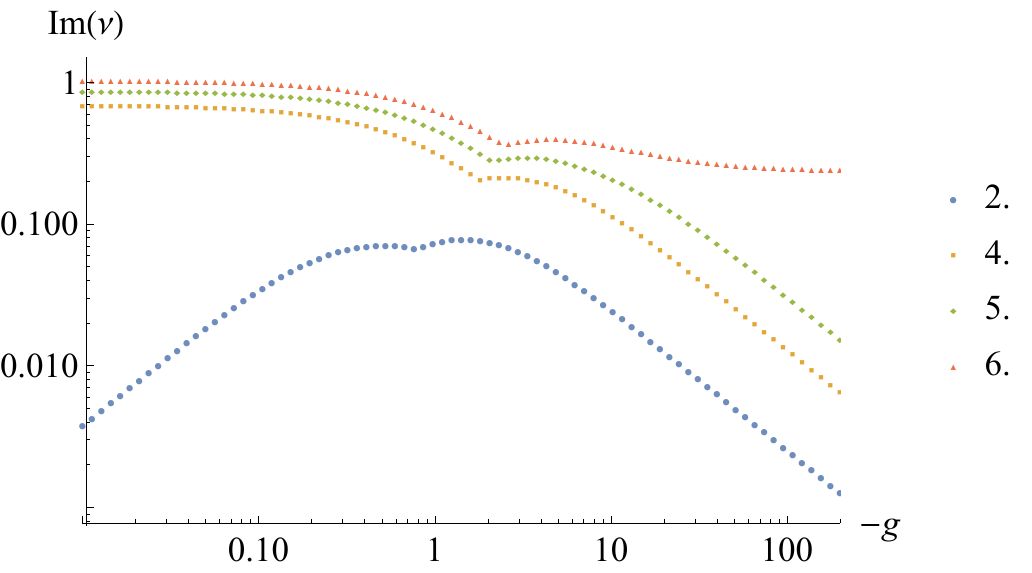}
\caption{\label{RatesVsGQ1}
	Growth rates versus damping gain at $q=1$; the legend shows values of $w$. The TMCI threshold $w=2.3$;  3FMCI wake threshold is between $w=5$ and $w=6$.}
\end{figure}
\begin{figure}[h]
\includegraphics[width=\linewidth]{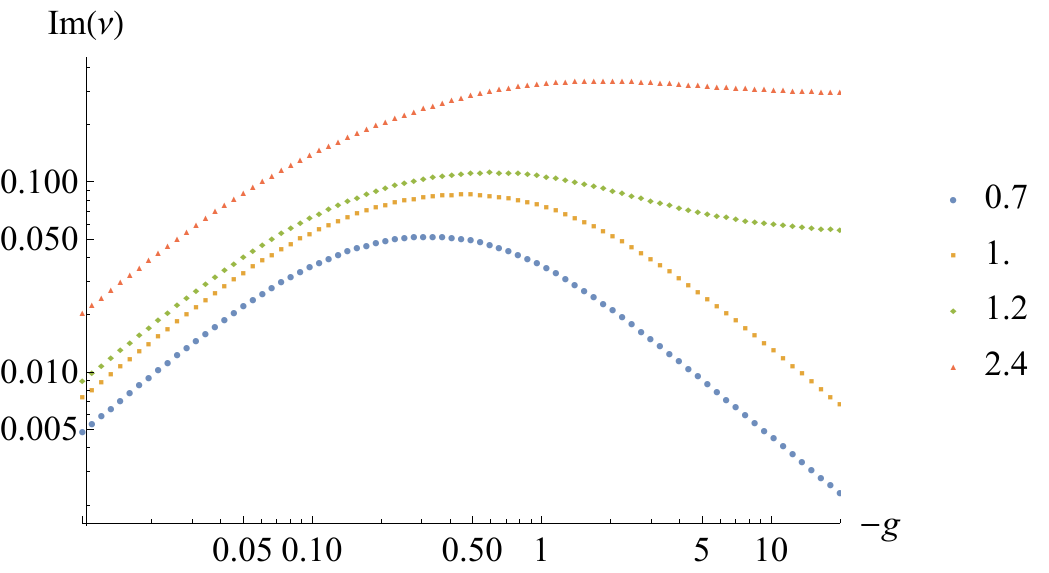}
\caption{\label{RatesVsGQ10}
	Same as Fig.~\ref{RatesVsGQ1}, for SC parameter $q=10$. TMCI vanished, 3FMCI threshold $w_\mathrm{3F}$ is shown to be $1<w_\mathrm{3F}<1.2$
	 }
\end{figure}
\begin{figure}[h!]
\includegraphics[width=\linewidth]{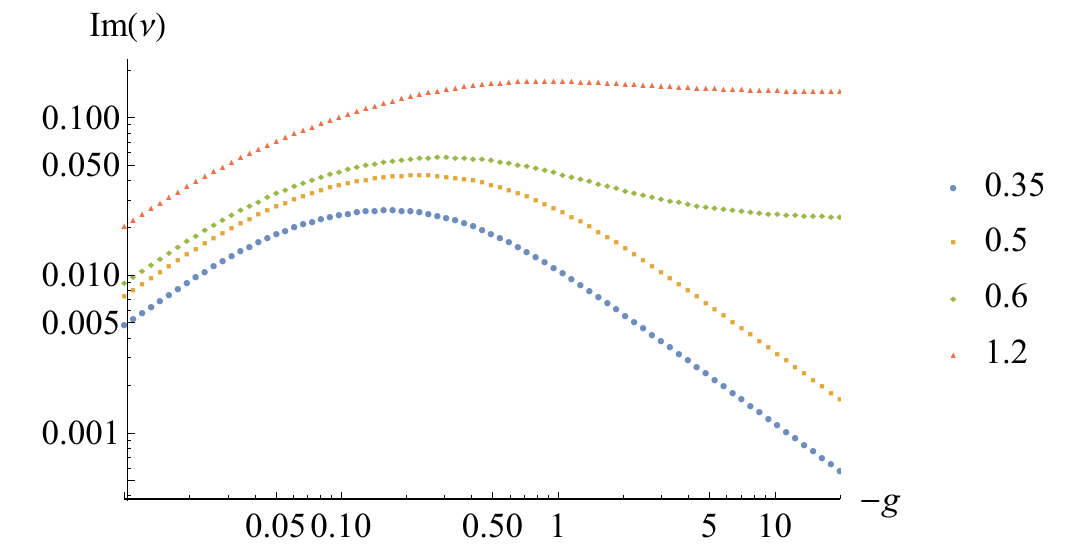}
\caption{\label{RatesVsGQ20}
	Same for $q=20$. The 3FMCI wake threshold is shown to be between $0.5$ and $0.6\,$. Comparison with Fig.~\ref{RatesVsGQ10} suggests a hypothesis for this threshold $\,w_\mathrm{3F} \simeq 10/q\,$  at $\,q \gg 1$. }
\end{figure}

Figures~\ref{RatesVsGQ1}~-~\ref{RatesVsGQ20} suggest to look at 3FMCI through the growth rate versus gain dependences at various wakes, as in Fig.~\ref{RatesVsGQ0}, but for non-zero SC. Figure~\ref{RatesVsGQ1} with its moderate SC, $q=1$, demonstrates the typical FWI pattern for $w=2$, i.e. slightly below TMCI threshold: the growth rate linearly increases with gain up to a certain its value, and then drops in the inverse proportion to it, when it is well above that value. Data points for $w=4$ and $w=5$ relate to TMCI of the modes $0$ and $-1$; they show a possibility of suppressing the instability with the sufficiently large gain, decoupling the coupled modes and turning it into FWI with its decreasing growth rate. Data points for $w=6$ show a case when that sort of instability suppression does not work. The reason is that for such a large wake, the 3FMCI scenario is realized: while the modes $0$ and $-1$ are decoupled at a sufficiently large gain, the positive modes $1$ and $2$ are coupled by the common action of wake, SC and feedback. Figures~\ref{RatesVsGQ10}~and~\ref{RatesVsGQ20} address the strong SC case, $q \gg 1$. While TMCI vanishes in this case, 3FMCI shows itself, with its wake threshold demonstrating inverse proportionality over SC, $w_\mathrm{3F} \simeq 10/q$. For all the 3FMCI cases of Figs.~\ref{RatesVsGQ1}~-~\ref{RatesVsGQ20} , the coupled modes are $+1^\mathrm{st}$ and $+2^\mathrm{nd}$.

\section{ Short Wakes} \label{Sec:ShortWake}

Up to this point, all the illustrations were performed for the theta-wake, $W(s)=\Theta(s)$.
In this section we make a brief excursion to the land of short wakes, or long bunches, to see what may change in comparison with long wakes, which our theta-wake model represented above.  
\begin{figure}[h]
\includegraphics[width=\linewidth]{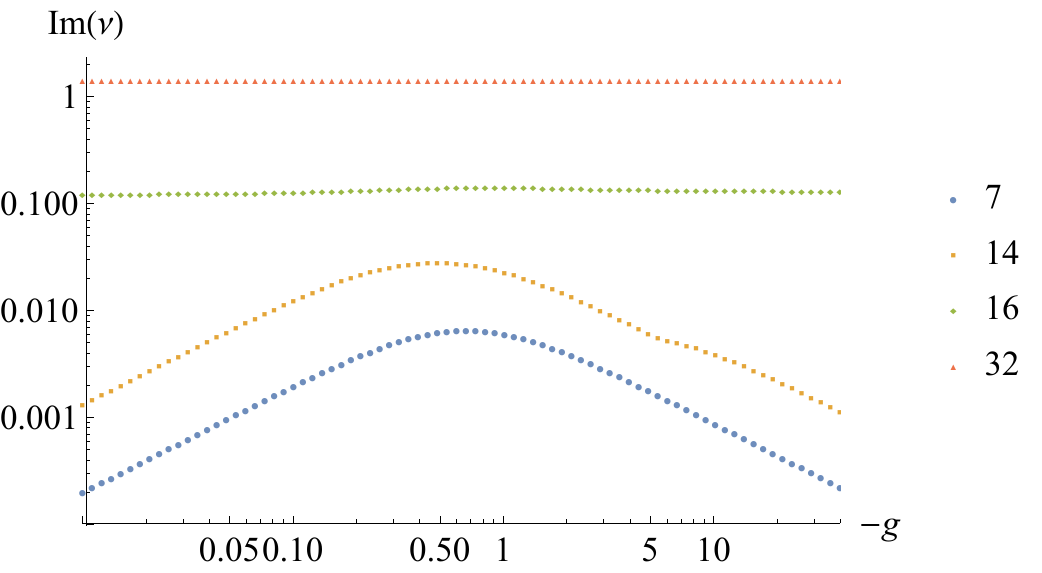}
\caption{\label{RatesVsGQ0kr10}
	Broadband wake with the phase advance $k_r=10$. No SC, $\,q=0\,$. TMCI wake threshold $w_\mathrm{th}^0=15$ is clearly seen.  }
\end{figure}
\begin{figure}[h]
\includegraphics[width=\linewidth]{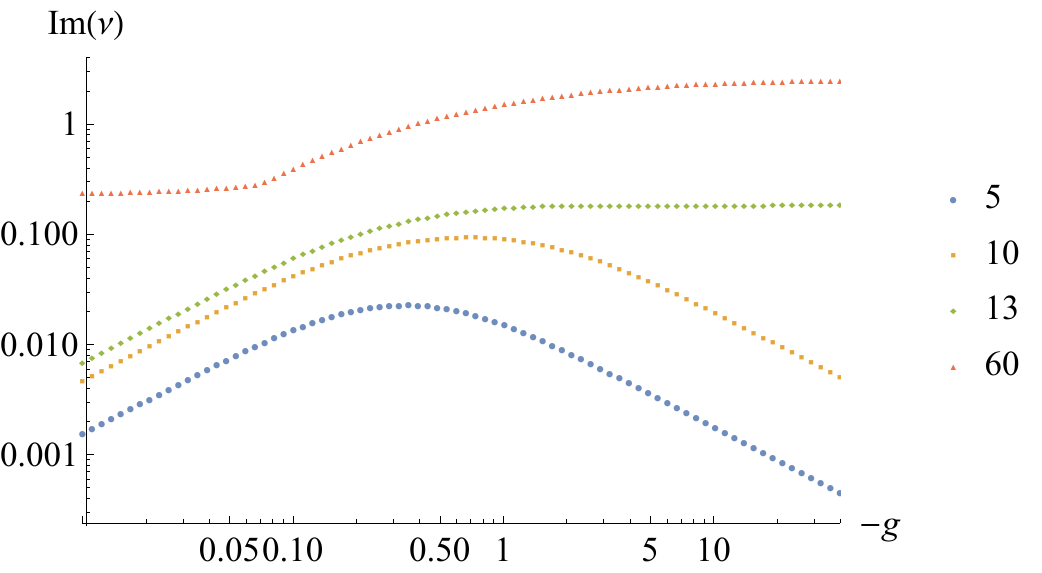}
\caption{\label{RatesVsGQ5kr10}
	Same as Fig.~\ref{RatesVsGQ0kr10}, but with SC $\,q=5\,$. TMCI threshold in this case $w_\mathrm{th}^0=48$. At $w \geq13\,$ and $g \geq 1\,$, the modes $+1^\mathrm{st}$ and $+2^\mathrm{nd}$ couple, resulting in 3FMCI. }
\end{figure}

As a representative of the short wakes, let us take a conventional broadband wake,
\begin{equation}
\begin{split}
& W(s) = \sin (\bar{k}s) \exp (-\alpha s)\,; \\
& \alpha \equiv k_r/(2 Q_r),\;\; \bar{k} \equiv \sqrt{k_r^2 - \alpha^2}\,,
\end{split}
\label{BBWake}
\end{equation}
with the quality factor $Q_r=1$ and a rather large phase advance $k_r=10$. For this wake, Figs.~\ref{RatesVsGQ0kr10}~and~\ref{RatesVsGQ5kr10} show what happens with the growth rates versus gain for various wakes. The former, Fig.~\ref{RatesVsGQ0kr10}, demonstrates no-SC case, with the TMCI threshold $w_\mathrm{th}^0=15$, clearly seen as a qualitative difference between $w=14$ and $w=16$ cases. Below the TMCI threshold, we see there a typical FWI single-maximum behavior of the growth rate, with its proportional and inversely proportional sides. Above the TMCI threshold, the growth rate is totally insensitive to the gain. It can be shown by means of the eigenfunction analysis similar to Fig.~\ref{GridXSCGCoupled} that the modes which couple here are $-1$ and $-2$. Thus, it is no surprise that the feedback does not affect TMCI in this case: these coupled modes do not contribute to the center of mass offset. 

What happens for the short wake with some considerable SC is presented in Fig.~\ref{RatesVsGQ5kr10} for $q=5$, where three different instabilities can be seen. First, at small wakes, $w \leq 10$, the typical FWI single-maximum pattern is seen. Then, at higher wake, $w \geq13$, 3FMCI takes place, as soon as the gain is sufficiently large, $g>1$. It can be shown that the 3FMCI coupled modes are $+1$ and $+2$, which should be expected, since the mode $+1$ is mostly connected with this wake, being also significantly head-to-tail amplified. The third instability recognizable in this plot is TMCI, which threshold can be computed as $w_\mathrm{th}^0=48$, so the case $w=60$ does indeed represent it, with its $-1^\mathrm{st}$ and $-2^\mathrm{nd}$ modes coupled. However, if the gain is not too small, $g \geq 0.05$, the most unstable mode is not one of the two negative modes, but the mode $+1$, which is absolute-convectively unstable (AC-unstable) at moderate gains and 3FMC-unstable at higher gains.

\section{ Summary}

In this paper, single-bunch instabilities were examined, with only a short-range wake, space charge and a resistive feedback taken into account. For the previously discussed {\bf f}eedback-and-{\bf w}ake {\bf i}nstability, {\sf FWI},~\cite{PhysRevSTAB.17.021007, PhysRevAccelBeams.19.084402, Metral:2018lds}, its general single-maximum dependence on the gain was demonstrated, with proportional and inversely proportional sides there. A new type of {\bf 3}-{\bf f}actor {\bf m}ode-{\bf c}oupling {\bf i}nstability, {\sf 3FMCI}, where positive modes couple due to a common action of {\bf f}eedback, Coulomb {\bf f}ield and wake {\bf f}ield, was demonstrated both for long and short wakes. 

The analysis was performed within the framework of the ABS model, as described in Sec.~\ref{Sec:ABS}. The approach of this paper, as well as of Refs.~\cite{Burov:2018pjl, Burov:2018rmx}, can be called mathematico-empirical: qualitative and structural features of solutions of a multi-parametric, multi-variable model are demonstrated on examples and discussed as reasonable expectations, instead of being proved by theorems. Certainly, theorems would be extremely valuable in this sort of analysis, but often we do not know what can be proven and almost never know how. That is why, trying to describe sets of solutions of a complicated model, we usually can do it only in this manner of zoological depictions and classifications.    

I am thankful to Elias Metral for pointing my attention to Ref.~\cite{Cappi:2000ze} and multiple clarifications given in the related discussion.

Fermilab is operated by Fermi Research Alliance, LLC under Contract No. DE-AC02-07CH11359 with the United States Department of Energy.



\bibliography{bibfile}			

\providecommand{\noopsort}[1]{}\providecommand{\singleletter}[1]{#1}%
\begin{thebibliography}{15}%
\makeatletter
\providecommand \@ifxundefined [1]{%
 \@ifx{#1\undefined}
}%
\providecommand \@ifnum [1]{%
 \ifnum #1\expandafter \@firstoftwo
 \else \expandafter \@secondoftwo
 \fi
}%
\providecommand \@ifx [1]{%
 \ifx #1\expandafter \@firstoftwo
 \else \expandafter \@secondoftwo
 \fi
}%
\providecommand \natexlab [1]{#1}%
\providecommand \enquote  [1]{``#1''}%
\providecommand \bibnamefont  [1]{#1}%
\providecommand \bibfnamefont [1]{#1}%
\providecommand \citenamefont [1]{#1}%
\providecommand \href@noop [0]{\@secondoftwo}%
\providecommand \href [0]{\begingroup \@sanitize@url \@href}%
\providecommand \@href[1]{\@@startlink{#1}\@@href}%
\providecommand \@@href[1]{\endgroup#1\@@endlink}%
\providecommand \@sanitize@url [0]{\catcode `\\12\catcode `\$12\catcode
  `\&12\catcode `\#12\catcode `\^12\catcode `\_12\catcode `\%12\relax}%
\providecommand \@@startlink[1]{}%
\providecommand \@@endlink[0]{}%
\providecommand \url  [0]{\begingroup\@sanitize@url \@url }%
\providecommand \@url [1]{\endgroup\@href {#1}{\urlprefix }}%
\providecommand \urlprefix  [0]{URL }%
\providecommand \Eprint [0]{\href }%
\providecommand \doibase [0]{http://dx.doi.org/}%
\providecommand \selectlanguage [0]{\@gobble}%
\providecommand \bibinfo  [0]{\@secondoftwo}%
\providecommand \bibfield  [0]{\@secondoftwo}%
\providecommand \translation [1]{[#1]}%
\providecommand \BibitemOpen [0]{}%
\providecommand \bibitemStop [0]{}%
\providecommand \bibitemNoStop [0]{.\EOS\space}%
\providecommand \EOS [0]{\spacefactor3000\relax}%
\providecommand \BibitemShut  [1]{\csname bibitem#1\endcsname}%
\let\auto@bib@innerbib\@empty
\bibitem [{\citenamefont {Blaskiewicz}(1998)}]{blaskiewicz1998fast}%
  \BibitemOpen
  \bibfield  {author} {\bibinfo {author} {\bibfnamefont {M.}~\bibnamefont
  {Blaskiewicz}},\ }\href@noop {} {\bibfield  {journal} {\bibinfo  {journal}
  {Physical Review Special Topics-Accelerators and Beams}\ }\textbf {\bibinfo
  {volume} {1}},\ \bibinfo {pages} {044201} (\bibinfo {year}
  {1998})}\BibitemShut {NoStop}%
\bibitem [{\citenamefont {Ng}\ and\ \citenamefont {Burov}(1999)}]{Ng:1999fy}%
  \BibitemOpen
  \bibfield  {author} {\bibinfo {author} {\bibfnamefont {K.~Y.}\ \bibnamefont
  {Ng}}\ and\ \bibinfo {author} {\bibfnamefont {A.~V.}\ \bibnamefont {Burov}},\
  }\bibfield  {booktitle} {\emph {\bibinfo {booktitle} {{Instabilities of high
  intensity hadron beams in rings. Proceedings, Workshop, Brookhaven, Upton,
  USA, June 28-July 1, 1999}}},\ }\href {\doibase 10.1063/1.1301875} {\bibfield
   {journal} {\bibinfo  {journal} {AIP Conf. Proc.}\ }\textbf {\bibinfo
  {volume} {496}},\ \bibinfo {pages} {49} (\bibinfo {year} {1999})}\BibitemShut
  {NoStop}%
\bibitem [{\citenamefont {Burov}(2009)}]{burov2009head}%
  \BibitemOpen
  \bibfield  {author} {\bibinfo {author} {\bibfnamefont {A.}~\bibnamefont
  {Burov}},\ }\href@noop {} {\bibfield  {journal} {\bibinfo  {journal}
  {Physical Review Special Topics-Accelerators and Beams}\ }\textbf {\bibinfo
  {volume} {12}},\ \bibinfo {pages} {044202} (\bibinfo {year}
  {2009})}\BibitemShut {NoStop}%
\bibitem [{\citenamefont {Balbekov}(2011)}]{PhysRevSTAB.14.094401}%
  \BibitemOpen
  \bibfield  {author} {\bibinfo {author} {\bibfnamefont {V.}~\bibnamefont
  {Balbekov}},\ }\href {\doibase 10.1103/PhysRevSTAB.14.094401} {\bibfield
  {journal} {\bibinfo  {journal} {Phys. Rev. ST Accel. Beams}\ }\textbf
  {\bibinfo {volume} {14}},\ \bibinfo {pages} {094401} (\bibinfo {year}
  {2011})}\BibitemShut {NoStop}%
\bibitem [{\citenamefont {Zolkin}\ \emph {et~al.}(2017)\citenamefont {Zolkin},
  \citenamefont {Burov},\ and\ \citenamefont {Pandey}}]{Zolkin:2017sdv}%
  \BibitemOpen
  \bibfield  {author} {\bibinfo {author} {\bibfnamefont {T.}~\bibnamefont
  {Zolkin}}, \bibinfo {author} {\bibfnamefont {A.}~\bibnamefont {Burov}}, \
  and\ \bibinfo {author} {\bibfnamefont {B.}~\bibnamefont {Pandey}},\
  }\href@noop {} {\  (\bibinfo {year} {2017})},\ \Eprint
  {http://arxiv.org/abs/1711.11110} {arXiv:1711.11110 [physics.acc-ph]}
  \BibitemShut {NoStop}%
\bibitem [{\citenamefont {Ruth}\ and\ \citenamefont
  {Wang}(1981)}]{Ruth:1981xt}%
  \BibitemOpen
  \bibfield  {author} {\bibinfo {author} {\bibfnamefont {R.~D.}\ \bibnamefont
  {Ruth}}\ and\ \bibinfo {author} {\bibfnamefont {J.~M.}\ \bibnamefont
  {Wang}},\ }\bibfield  {booktitle} {\emph {\bibinfo {booktitle} {{APPLICATION
  OF ACCELERATORS IN RESEARCH AND INDUSTRY. PROCEEDINGS, 6TH CONFERENCE,
  DENTON, TEXAS, USA, NOVEMBER 3-5, 1980}}},\ }\href {\doibase
  10.1109/TNS.1981.4331707} {\bibfield  {journal} {\bibinfo  {journal} {IEEE
  Trans. Nucl. Sci.}\ }\textbf {\bibinfo {volume} {28}},\ \bibinfo {pages}
  {2405} (\bibinfo {year} {1981})}\BibitemShut {NoStop}%
\bibitem [{\citenamefont {Burov}\ and\ \citenamefont
  {Lebedev}(2009)}]{PhysRevSTAB.12.034201}%
  \BibitemOpen
  \bibfield  {author} {\bibinfo {author} {\bibfnamefont {A.}~\bibnamefont
  {Burov}}\ and\ \bibinfo {author} {\bibfnamefont {V.}~\bibnamefont
  {Lebedev}},\ }\href {\doibase 10.1103/PhysRevSTAB.12.034201} {\bibfield
  {journal} {\bibinfo  {journal} {Phys. Rev. ST Accel. Beams}\ }\textbf
  {\bibinfo {volume} {12}},\ \bibinfo {pages} {034201} (\bibinfo {year}
  {2009})}\BibitemShut {NoStop}%
\bibitem [{\citenamefont {Cappi}\ \emph {et~al.}(2000)\citenamefont {Cappi},
  \citenamefont {Metral},\ and\ \citenamefont {Metral}}]{Cappi:2000ze}%
  \BibitemOpen
  \bibfield  {author} {\bibinfo {author} {\bibfnamefont {R.}~\bibnamefont
  {Cappi}}, \bibinfo {author} {\bibfnamefont {E.}~\bibnamefont {Metral}}, \
  and\ \bibinfo {author} {\bibfnamefont {G.}~\bibnamefont {Metral}},\ }in\
  \href {http://weblib.cern.ch/abstract?CERN-PS-2000-017-AE} {\emph {\bibinfo
  {booktitle} {{Particle accelerator. Proceedings, 7th European Conference,
  EPAC 2000, Vienna, Austria, June 26-30, 2000. Vol. 1-3}}}}\ (\bibinfo {year}
  {2000})\ pp.\ \bibinfo {pages} {1152--1154}\BibitemShut {NoStop}%
\bibitem [{\citenamefont {Burov}(2018{\natexlab{a}})}]{Burov:2018pjl}%
  \BibitemOpen
  \bibfield  {author} {\bibinfo {author} {\bibfnamefont {A.}~\bibnamefont
  {Burov}},\ }\href@noop {} {\  (\bibinfo {year} {2018}{\natexlab{a}})},\
  \Eprint {http://arxiv.org/abs/1807.04887} {arXiv:1807.04887 [physics.acc-ph]}
  \BibitemShut {NoStop}%
\bibitem [{\citenamefont {Burov}(2018{\natexlab{b}})}]{Burov:2018rmx}%
  \BibitemOpen
  \bibfield  {author} {\bibinfo {author} {\bibfnamefont {A.}~\bibnamefont
  {Burov}},\ }\href@noop {} {\  (\bibinfo {year} {2018}{\natexlab{b}})},\
  \Eprint {http://arxiv.org/abs/1808.08498} {arXiv:1808.08498 [physics.acc-ph]}
  \BibitemShut {NoStop}%
\bibitem [{\citenamefont {Chao}(1993)}]{chao1993physics}%
  \BibitemOpen
  \bibfield  {author} {\bibinfo {author} {\bibfnamefont {A.~W.}\ \bibnamefont
  {Chao}},\ }\href@noop {} {\emph {\bibinfo {title} {Physics of collective beam
  instabilities in high energy accelerators}}}\ (\bibinfo  {publisher}
  {Wiley},\ \bibinfo {year} {1993})\BibitemShut {NoStop}%
\bibitem [{\citenamefont {Burov}(2014)}]{PhysRevSTAB.17.021007}%
  \BibitemOpen
  \bibfield  {author} {\bibinfo {author} {\bibfnamefont {A.}~\bibnamefont
  {Burov}},\ }\href {\doibase 10.1103/PhysRevSTAB.17.021007} {\bibfield
  {journal} {\bibinfo  {journal} {Phys. Rev. ST Accel. Beams}\ }\textbf
  {\bibinfo {volume} {17}},\ \bibinfo {pages} {021007} (\bibinfo {year}
  {2014})}\BibitemShut {NoStop}%
\bibitem [{\citenamefont
  {Burov}(2016{\natexlab{a}})}]{PhysRevAccelBeams.19.084402}%
  \BibitemOpen
  \bibfield  {author} {\bibinfo {author} {\bibfnamefont {A.}~\bibnamefont
  {Burov}},\ }\href {\doibase 10.1103/PhysRevAccelBeams.19.084402} {\bibfield
  {journal} {\bibinfo  {journal} {Phys. Rev. Accel. Beams}\ }\textbf {\bibinfo
  {volume} {19}},\ \bibinfo {pages} {084402} (\bibinfo {year}
  {2016}{\natexlab{a}})}\BibitemShut {NoStop}%
\bibitem [{\citenamefont {Metral}\ \emph {et~al.}(2018)\citenamefont {Metral},
  \citenamefont {Amorim}, \citenamefont {Antipov}, \citenamefont {Biancacci},
  \citenamefont {Buffat},\ and\ \citenamefont {Li}}]{Metral:2018lds}%
  \BibitemOpen
  \bibfield  {author} {\bibinfo {author} {\bibfnamefont {E.}~\bibnamefont
  {Metral}}, \bibinfo {author} {\bibfnamefont {D.}~\bibnamefont {Amorim}},
  \bibinfo {author} {\bibfnamefont {S.}~\bibnamefont {Antipov}}, \bibinfo
  {author} {\bibfnamefont {N.}~\bibnamefont {Biancacci}}, \bibinfo {author}
  {\bibfnamefont {X.}~\bibnamefont {Buffat}}, \ and\ \bibinfo {author}
  {\bibfnamefont {K.}~\bibnamefont {Li}},\ }in\ \href {\doibase
  10.18429/JACoW-IPAC2018-THPAF048} {\emph {\bibinfo {booktitle} {{Proceedings,
  9th International Particle Accelerator Conference (IPAC 2018): Vancouver, BC
  Canada}}}}\ (\bibinfo {year} {2018})\ p.\ \bibinfo {pages}
  {THPAF048}\BibitemShut {NoStop}%
\bibitem [{\citenamefont {Burov}(2016{\natexlab{b}})}]{Burov:2016jsh}%
  \BibitemOpen
  \bibfield  {author} {\bibinfo {author} {\bibfnamefont {A.}~\bibnamefont
  {Burov}},\ }\href@noop {} {\  (\bibinfo {year} {2016}{\natexlab{b}})},\
  \Eprint {http://arxiv.org/abs/1606.07430} {arXiv:1606.07430 [physics.acc-ph]}
  \BibitemShut {NoStop}%
\end{thebibliography}%

\end{document}